\documentclass[11pt]{article}
\usepackage{amssymb}
\begin{document}

\rightline{FT-UAM-06-3}
\rightline{IFT-UAM/CSIC-06-16}
\rightline{FFUOV-06/01}
\rightline{hep-th/0604031}
\rightline{April 2006}
\vspace{2truecm}

\centerline{\LARGE \bf Computing Wilson lines with dielectric
  branes} 
\vspace{1.3truecm}

\centerline{
    {\large \bf D. Rodr\'{\i}guez-G\'omez}
                                                       }

\vspace{.4cm}
\centerline{{\it Departamento de F{\'\i}sica}}
\centerline{{\it Universidad de Oviedo}}
\centerline{{\it Avda.~Calvo Sotelo 18, 33007 Oviedo, Spain}}
\centerline{{\it and}}
\centerline{{\it Departamento de F{\'\i}sica Te{\'o}rica C-XI}}
\centerline{{\it Universidad Aut{\'o}noma de Madrid}}
\centerline{{\it Cantoblanco, 28049, Madrid, Spain}}
\vspace{0.2cm}
\centerline{\tt Diego.Rodriguez.Gomez@cern.ch}

\vspace{2truecm}
                                       
\centerline{\bf ABSTRACT}
\vspace{.5truecm}

\noindent
Wilson lines in $\mathcal{N}=4$ SYM can be computed in terms of branes carrying electric flux, \textit{i.e.} F-strings dissolved in their
worldvolumes. It is then natural to think that those configurations are the
effective description of strings expanding due to dielectric effect to D-branes. In this note we explicitly show this for a class of such
configurations, namely those dual to Wilson lines either in the symmetric or
in the antisymmetric tensor product of fundamentals.

\newpage
\section{Introduction}

The $AdS/CFT$ correspondence \cite{malda1} relates gravity theories in $AdS$
spaces with certain field theories. In particular, it identifies type IIB
string theory on $AdS_5\times S^5$ with $\mathcal{N}=4$ SYM.

In the field theory side of the correspondence, a very interesting operator is
the Wilson loop. Recently, very much progress has been done in understanding
these kind of operators, both from the string theory side and the CFT side
(\textit{e.g.} \cite{ESZ},\cite{GD}\dots). 

In \cite{malda2} and \cite{RY} a specific proposal to compute Wilson loops in
terms of the dual string theory was put forward. To be precise, the strong
coupling version of the Wilson loop is computed by means of the area of the
worldsheet of a string which, at the boundary of $AdS$, terminates on the
loop. However, with these techniques, one cannot obtain the subleading
corrections which arise when one considers coincident Wilson loops, multiply
wound Wilson loops or Wilson loops in a higher dimensional representation
\cite{DS}. In a beautiful paper, Drukker and Fiol \cite{DF} showed that it is
possible to compute a class of these loops (for example those with many
coincident Wilson loops) using D branes carrying a large fundamental string
charge dissolved on their worldvolume pinching off at the boundary of the
$AdS$ on the Wilson loop. The idea behind this proposal is that the D-brane
can be seen as the effective description of a multi-string configuration when
the number of strings is large, in very much of the spirit of \cite{callan},
\cite{emparan} and \cite{myers}. Thus, the D3 brane represents a way of taking
into account all the interactions among the strings. 

Similar configurations but with D5 branes carrying electric DBI flux were
recently considered in \cite{HK} and \cite{Y2}\footnote{Let us also point that
  this type of branes has been considered in the literature for other
  purposes (\cite{Ramallo}).}. In the first reference, a D5
brane in the euclidian Schwarzschild black hole in $AdS$ carrying electric
flux was proposed as dual to the thermal Polyakov loop. One can see that there
are several possible configurations for such a brane. The extremal version of
one of them is proposed in the second reference as the dual of a Wilson loop
on an antisymmetric tensor product of the fundamental representation. In
\cite{Gomis} more work along this line was done, showing how the
group-theoretic structure of the associated Wilson line arises. The final
picture for this Wilson line coming from those papers is that the
infinite straight Wilson line in a symmetric tensor product of fundamentals
can be described by means of a single D3 brane carrying F1 strings dissolved,
while the infinite straight Wilson line in an antisymmetric tensor product of
fundamentals can be computed by means of a D5 brane with fundamental strings
on it. Clearly, both branes should pinch off on the Wilson line at the boundary
in order to capture the desired field theoretic configuration. In some sense,
the brane duals arising from this picture are in great parallelism to the
giant graviton phenomenon of \cite{giants}, to which also a microscopic
description along the lines presented in this note was given in a series of
papers \cite{micdesc}.

In this note we will concentrate on the straight Wilson line,
which preserve one half of the total supersymmetries. We will explicitly show
how this branes should be seen as the effective description of a collective of
strings pinching off on the loop at the boundary (and thus following the
original recipe to compute the Wilson loop in the $AdS/CFT$ context). Once the
interactions between them are taken into account, due to dielectric effect
(\cite{myers}), they undergo an expansion into higher dimensional objects
which, in the adequate limit, are effectively described with the so far
mentioned branes. In a sense, this properly justifies why the these branes
capture the desired Wilson loops, since they are nothing but the effective
description of a collection of strings following the prescription in
\cite{malda2} and \cite{RY}.

In the case of the symmetric product representation Wilson line, since
fundamental strings naturally expand to D3 branes; one should expect to find a
microscopical description in terms of dielectric fundamental strings expanding
to a D3 brane, in such a way that once a suitable limit is taken, the
macroscopic D3 of \cite{DF} is recovered as the effective description. Making
use of the S-duality invariance of the background, we will actually compute
the expansion of coincident D1 branes into a D3.  Thus we will use the
action for dielectric D1 branes of \cite{myers}. Then, due to the S-duality
invariance of the background, the functional form of the action for such a
configuration will be the same as the one we are interested in. In any case, we
should be aware that strictly speaking we are computing a 't Hooft line. We
will explicitly show this, along the lines of \cite{DF}. However, there is a
way of computing directly the Wilson loop, namely by constructing an action for
dielectric fundamental strings. This action can be constructed
(\cite{bariones}) by means of a chain of dualities starting with the action
for dielectric gravitons in M-theory presented in
\cite{dielectricgravitons}. However, as it should be expected, once we
particularize it for our background it reduces exactly to Myers action, thus
giving the same functional form for both the Wilson and 't Hooft line. 

The case of the D5 is more subtle, since naturally F1 strings do not polarize
into D5 branes, but into NS5 branes. The natural candidate to find a
microscopical description for the D5 is the D1 brane. We will see explicitly
how we can indeed achieve a microscopical description of the D5 brane
configurations in \cite{HK}, \cite{Y2}, \cite{Gomis} in terms of dielectric D1
branes. In turn, we will have to add the F1 charge as an electric DBI vector on
the worldvolume of such a dielectric configuration. In addition, in order to
capture the desired macroscopic configurations, we should restrict to those
configurations in which the net D1 brane charge is zero. Furthermore, inspired
by the latter construction in which we have at the same time both D1 and F1
charge pinching off on the line at the boundary, an interesting possibility arises, namely that of computing Wilson-'t
Hooft lines. We will see that if we do not remove the extra D1 charge what we
get is a configuration in which what pinches off on the boundary is not F1
charge but a dyonic string, carrying also D1 brane charge. This should be
capturing information about the associated Wilson-'t Hooft line. Indeed, a
similar computation can be done in the symmetric product representation Wilson
line, in terms of a D3 brane which carries both F1 and D1 charge, which should
also contain information about the Wilson-'t Hooft line in a symmetric product
representation.

 It is worth to mention that since we will be interested in the blowing
up of strings to D-branes, we will focus in the bulk computation. The boundary
terms to add in order to have the suitable boundary conditions will be exactly
as those in the literature, and we refer the reader to the corresponding
papers.

\section{Infinite straight Wilson line in the symmetric tensor product
  representation}

In this section we will use as coordinates for $AdS_5$

\begin{equation}
ds^2=\frac{L^2}{y^2}\big(dy^2+d\vec{x}^2\big)\ .
\end{equation}

We will take the infinite Wilson line in the symmetric tensor product of
fundamental representation to be extended along $x^1$. Clearly, there is
an $SO(3)$ rotational invariance around this line. 

As it is well-known, for this Wilson line $\langle W\rangle=1$ (see
\cite{ESZ}), and, following \cite{DF}, we would like to see this rather than
with a single fundamental string ending on the line at the boundary, with a D3
brane carrying electric flux on it representing $M$ strings and pinching off
at the boundary on the line. In order to do this, it is convenient to switch
to the following coordinates

\begin{equation}
ds^2=\frac{L^2}{y^2}\big(dy^2+dt^2+dr^2+r^2d\Omega_2^2\big)\ .
\end{equation}

\noindent In addition, there is a 4-form RR potential given by

\begin{equation}
C^{(4)}=\frac{L^4r^2}{y^4}\ dt\wedge dr\wedge d\omega_2\ ,
\end{equation}

\noindent where $d\omega_2$ stands for the volume form on the $S^2$.

Because of the symmetries of the problem, it is clear that we should consider
as dual of the infinite Wilson line a D3 brane whose worldvolume coordinates
are $\{t,r,\Omega_2\}$, and take $y=y(r)$. As one can see in \cite{DF}, this
D3 brane pinches off at the boundary over the Wilson line along $t$. 

In addition, we have to add electric DBI flux on the brane representing the
strings. Therefore we will take a non-zero electric DBI field $F_{tr}=F$. For
such a D brane the action is easily seen to be

\begin{equation}
S=\int dt dr\
2T_1\frac{L^4r^2}{y^4}\Big(\sqrt{1+y'^2-\frac{y^4}{L^4}F^2}-1\Big)\ ,
\end{equation}

\noindent where, following \cite{HK}, we have taken $F\rightarrow iF$ in order
to properly represent F strings.

For future convenience, we re-write everything in terms of the 1-brane tension
$T_1$. In addition, we are taking units so that $2\pi l_s^2=1$.

The prime represents derivative with respect to $r$.

Since in the action the DBI potential does not explicitly appear in the
action, we can change the DBI field strength for its conserved momentum $P$ to
obtain a ``routhian''. 

\begin{equation}
\mathcal{R}_{Macro}=\int dt dr\ 2T_1\frac{L^4r^2}{y^4}\Big(\sqrt{1+y'^2}\sqrt{1+\frac{y^4P^2}{4r^4T_1^2L^4}}-1\Big)\ ,
\end{equation}

Here the conserved momentum represents the number of dissolved strings, so, in
units of $T_1$, it will be an integer $M$.

Although it is not the main aim of this work, let us mention that, once the
appropriate boundary terms are added, in order to ensure the right boundary
conditions, the resulting action leads to an equation of motion which has as a
solution $y\sim r$. This clearly displays the fact that the brane pinches off
on the line, while once we evaluate the action of the brane in the solution we
get that $S=0$, which leads to $\langle W\rangle =1$ (\cite{ESZ}, \cite{DF}).

The configuration above was engineered in such a way
that it captures the behavior of a large number of fundamental strings. In
view of the dielectric effect, it then is natural to search
for a dielectric configuration of strings expanding to the D3 brane
configuration. Since our background is S-duality invariant, we will describe
microscopically our configurations using the action for coincident D1 strings
\cite{myers}. We will take the strings along the coordinates $\{t,r\}$, and
assume that they expand to the transverse $S^2$. In addition, we will consider
a worldvolume scalar field $y=y(r)$. In cartesian coordinates, the
background reads

\begin{equation}
ds^2=\frac{L^2}{y^2}\big(dy^2+dt^2+dr^2+r^2d\vec{x}^2\big)\ ;
\end{equation}

\noindent and

\begin{equation}
C^{(4)}=\frac{L^4r^2}{y^4}x^k\epsilon_{ijk}\ dt\wedge dr\wedge dx^i\wedge
dx^j\ ,
\end{equation}

\noindent where $\epsilon_{ijk}$ is the completely antisymmetric tensor, and
$i,j,k=i,2,3$.

As non-commutative ansatz we will take

\begin{equation}
\label{S2anstazt}
X^i=\frac{1}{\sqrt{C_2}}J^i\ ,
\end{equation}

\noindent where the $J^i$ are the generators of and $M$ dimensional
representation of $SU(2)$ whose Casimir is $C_2$. Clearly, this way it is
satisfied that $\vec{X}^2=1$ as a matrix identity.

Upon particularizing Myers action to our particular set up, after taking the
trace, we arrive to the following action

\begin{equation}
\mathcal{R}_{micro}=\int dt dr\ 2T_1\frac{L^4r^2}{y^4}\Big(\sqrt{1+y'^2}\sqrt{\frac{M^2}{C_2}+\frac{y^4M^2}{4r^4L^4}}-1\Big)\ ,
\end{equation}

In order to arrive to this expression we have made use of the properties of
the symmetrized trace, which we have computed up to order
$\mathcal{O}(C_2^{-2})$, since we are interested in comparing with the
effective macroscopical computation. Clearly, once we take the Myers limit of
large $M$, since $\frac{M}{\sqrt{C_2}}\sim 1$, both the expression for the
macroscopic and microscopic configurations match perfectly.

The microscopical computation was done in terms of dielectric D1
strings. However, we arrive to a perfect matching with the macroscopical
computation, which carries F string charge due to the S-duality invariance of
the configuration (namely, because the dilaton and the 2-forms are
zero, and the D3 is self S-dual). Strictly speaking we are computing a 't
Hooft line. Following \cite{DF}, let us do the proper dual macroscopical
description dual to the 't Hooft line, which will be now in terms of a D3 with
magnetic DBI field on it simulating $M$ units of D1 string charge to the
brane. In the CS action we would have

\begin{equation}
S_{CS}=T_3\int F\int dt dr\ (C^{(2)}+C^{(0)}B_{NS})_{xr}\ .
\end{equation}

\noindent Therefore it is clear that we should take
$F=\frac{M}{2}d\omega_2$. Then, after a straightforward computation, it is
easy to see that the action for such a brane is simply

\begin{equation}
\label{tHoft}
\mathcal{R}_{'t\ Hooft}=\int dt dr\
2T_1\frac{L^4r^2}{y^4}\Big(\sqrt{1+y'^2}\sqrt{1+\frac{y^4M^2}{4r^4
    L^4}}-1\Big)\ .
\end{equation}

\noindent This expression is directly the same as the Legendre transformed of
the D3 with dissolved F1, and in turn the same expression as the action for
the dielectric D1 strings, as expected due to the referred S-duality
invariance of the considered background.

It is possible to directly describe the Wilson loop by means of an action for
dielectric fundamental strings (which naturally expand also to a D3
brane). This action can be constructed (\cite{bariones}) by means of a
chain of dualities starting with the action for coincident gravitons in
M-theory in \cite{dielectricgravitons}, and, once particularized for the case
at hand, it coincides with Myers action, thus giving the same result. This is
in agreement with what has been described here following \cite{DF}, and is to
be expected due to the S-duality invariance of this D3-brane configuration.

\section{Polyakov lines and infinite straight Wilson lines in the
  antisymmetric product representation}

The Polyakov loop, defined as the $SU(N)$ holonomy around the Euclidean time,
is a very useful quantity in trying to determine the order of the phase
transition between strong and weak coupling of the $\mathcal{N}=4$ SYM theory
on $S^3\times S^1$ (see \textit{e.g.} \cite{Shenker}). In \cite{HK} it was
proposed that the multiply wound Polyakov loop can be calculated along the
lines of \cite{DF}, \textit{i.e.} using D-branes with large dissolved string
charge. Following the same logic as before, since these configurations are also
designed to pinch off on the Wilson loop while carrying string charge, it
should be possible to give a microscopical description of them in terms of the
dielectric effect. 

Since we are analyzing thermal loops, we will work with the euclidian
Schwarzschild black hole in $AdS$, given by the metric

\begin{equation}
\label{BHAdS}
ds^2=R^2fdt^2+\frac{R^2}{f}dr^2+R^2r^2\big(d\alpha^2+\sin^2\alpha
d\Omega_2^2\big)+R^2d\Omega_5^ 2\ ,
\end{equation}

\noindent where

\begin{equation}
f=1-\frac{r_+^2(1+r_+^2)}{r^2}+r^2\ .
\end{equation}

The extremal limit corresponds to $r_+\rightarrow 0$, which reduces to
$AdS_5\times S^5$. 

In \cite{HK} it was realized that the suitable brane to describe such a
configuration was not a 3-brane but a 5-brane. In \cite{Y2} and \cite{Gomis},
it was further proposed that this D5-brane, which pinches off on the boundary
and wraps an $S^4$ inside the $S^5$ of the background, indeed represents the
infinite straight Wilson line in an antisymmetric tensor product of
fundamentals. Evidence for this in terms of the group-theoretic structure of
the associated loop was given in \cite{Gomis}, which also confirms the analogy
between Wilson line operators and giant gravitons in the sense that both
correspond either to symmetric or antisymmetric product representations of half
BPS operators which in the gravity side can be described as branes carrying
some dissolved charge. Exactly as in the symmetric tensor product, we will see
that these branes are to be seen as the effective description of a
multi-string configuration pinching off on the loop at the boundary, following
the classical recipe for computing the Wilson loop. In the
appendix we also consider the D3-brane configuration of \cite{HK}, which was
shown no to be the right candidate for this configuration. However, it should
also be regarded as the effective description of a multi-F1 configuration.

\subsection{D5 branes}

Let us consider the D5 brane configurations of \cite{HK} and \cite{Y2}
(indeed, they have been already considered in the literature previously
\cite{Ramallo}\footnote{Indeed, in \cite{pedro}, a microscopical description is
  given once we expand the Myers DBI action in a power series. As we will see,
  the full action has specialties (some instanton number in particular) which
  are fundamental when describing this Wilson loop}). In suitable coordinates the background reads

\begin{equation}
ds^2=R^2fdt^2+\frac{R^2}{f}dr^2+R^2r^2d\omega_3^2+R^2d\gamma^2+R^2\sin^2\gamma
d\vec{x}^2\ ,
\end{equation}

\noindent where $\vec{x}$ are cartesian coordinates for the $S^4$, and satisfy
$\vec{x}^2=1$.

The relevant piece of the RR potential is

\begin{equation}
C^{(4)}_{ijkl}=R^4D(\gamma)x^m\epsilon_{ijklm}\ ;
\end{equation}

\noindent where

\begin{equation}
D(\gamma)=\sin^3\gamma
\cos\gamma+\frac{3}{2}\cos\gamma\sin\gamma-\frac{3(\gamma-\pi)}{2}\ .
\end{equation}

Since the natural candidates to find a microscopical description of the D5 are
now D1 strings, let us consider D1-branes with worldvolume coordinates
$\{t,r\}$ and take $\gamma=\gamma (r)$. In addition, let us add a DBI field on
the strings $F_{tr}=F$, which stands for the presence of the fundamental
strings.

We will take the ansatz that our D1 strings expand to the transverse $S^4$,
which will be a fuzzy 4-sphere of radius 1. We will closely follow the
construction of the fuzzy 4-sphere of \cite{CLT}. 

\subsubsection{The fuzzy $S^4$}

Without going into very much detail, let us simply point out that in order to
have a fuzzy $S^4$ we have to find 5 matrices such that $\vec{X}^2=1$ as a
matrix identity. We can achieve such matrices by taking the $n$-fold symmetric
tensor product of the $\Gamma$ matrices of the Clifford algebra underlying the
$SO(5)$ group, \textit{i.e.}

\begin{equation}
G_i^{(n)}=\big(\Gamma_i\otimes 1\otimes \cdots 1+1\otimes \Gamma_i\otimes
1\cdots \otimes 1+\cdots 1\otimes 1\otimes \Gamma_i\big)_{Sym}\ .
\end{equation}

\noindent These matrices satisfy the property that $\Sigma_i
(G_i^{(n)})^2=n(n+4)\times 1$. 

The dimension of the representation of the fuzzy $S^4$ can be seen to be

\begin{equation}
Tr(1)=N=\frac{(n+1)(n+2)(n+3)}{6}\ .
\end{equation}

If we call $c=n(n+4)$, it is then clear that we can construct our unit fuzzy
$S^4$ as

\begin{equation}
X^i=\frac{1}{\sqrt{c}}G_i^{(n)}\ .
\end{equation}

A useful property is then 

\begin{equation}
\epsilon_{ijklm}X^iX^kX^lX^m=\frac{8n+16}{n}\times 1\ .
\end{equation}

In addition, we define 

\begin{equation}
G_{ij}=\frac{1}{2}[G_i,G_j]\ ,
\end{equation}

\noindent which will be useful when computing the action of our
configuration. These $G_{ij}$ satisfy the properties stated in the appendix A
of \cite{CLT}.

For further details on the construction of the fuzzy $S^4$, we refer the
reader to \cite{CLT}.

\vspace{0.5cm}

Once we make use of the fuzzy $S^4$ construction, after a straightforward
computation, we arrive to the following action for our system, which in the
large $n$ limit in which we are interested in order to compare with the
macroscopical description reads

\begin{equation}
\label{SD5microFULL}
S=\int dt dr\
T_1R^2(\frac{n^3}{6}+\frac{2n}{3}R^4\sin^4\gamma)\sqrt{1+f\gamma'^2-\frac{F^2}{R^4}}-\int dt dr\ T_1\frac{2n}{3}R^4D(\gamma)F\ .
\end{equation}

Again, the prime denotes derivative with respect to $r$, and following
\cite{HK} we have taken $F\rightarrow iF$.

As we will see, our construction is very similar to that in \cite{CP2} in what
microscopically we are capturing an extra charge of the brane, which
macroscopically will show up as dissolved instantons in the
worldvolume. In that reference, a microscopical description of the M5 giant
graviton in $AdS_4\times S^7$ was found in terms of dielectric gravitons
expanding to a fuzzy $S^5$, which was constructed as a fibration of an $S^1$
over a fuzzy $CP^2$. The giant graviton carries momentum along a cycle in the
$S^7$. However, in the microscopical description although the action used
already assumes momentum along one direction, the computation captures
momentum also along the isometry of the fibering. This can be seen by
comparing with the action for a wrapped macroscopical M5, which carries
momentum along both the 11th direction (as instantons) and along an extra
direction. Since the configuration of giant graviton carries momentum in just
one direction, by comparing with the macroscopical description one learns how
to unplug this extra charge. In our situation a similar situation occurs. In
the case at hand removing this extra charge (also instantons from the
macroscopical point of view) amounts, as we will see, to keep just

\begin{equation}
\label{SD5micro}
S=n\Big\{ \int dt dr\ \frac{2T_1R^6}{3}\Big( \sin^4\gamma\sqrt{1+f\gamma'^2-\frac{F^2}{R^4}}-\frac{D(\gamma)F}{R^2}\Big)\Big\}\ .
\end{equation}

Let us note what we recover is not the action for a single 5 brane (see next
 subsection), but the action for $n$ of them. This means that we are actually
 describing the coincident D5 brane configurations in \cite{Gomis}

\subsubsection{Macroscopical D5 with instantons}

In order to see the macroscopical counterpart of the above microscopical
configuration, instead of directly consider those branes on \cite{HK}, we will
also add a magnetic DBI field on the $S^4$ satisfying that $F_{mag}=\star
F_{mag}$, where the star is taken with respect to the unit $S^4$ metric. The
existence of such an instanton on $S^4$ can be seen for example in
\cite{myers2}.      

It is straightforward to see that for a D5 with worlvolume coordinates
$\{t,r,\Omega_4\}$ and $\gamma=\gamma (r)$ with both electric ($F_{tr}$) and
magnetic (lets call it $B$, satisfying as mentioned $B=\star B$), the action
reads

\begin{equation}
\label{SD5macroFULL}
S=\int dt dr\
\frac{2T_1}{3}R^2(B\wedge B+R^4\sin^4\gamma)\sqrt{1+f\gamma'^2-\frac{F^2}{R^4}}-\int dt dr\ \frac{2T_1}{3}R^4D(\gamma)F\ .
\end{equation}

Note that (\ref{SD5macroFULL}) is exactly of the form of
(\ref{SD5microFULL}). From here we see that the first term in the latter
corresponds with the instanton charge in the macroscopic version. In order to
describe the configurations in \cite{HK} we have to eliminate those
instantons, which leads us to the following macroscopic configuration

\begin{equation}
\label{SD5macro}
S=\int dt dr\
\frac{2T_1}{3}R^6\Big(\sin^4\gamma\sqrt{1+f\gamma'^2-\frac{F^2}{R^4}}-
\frac{D(\gamma)F}{R^2}\Big)\ ,
\end{equation}

\noindent which matches perfectly with that of (\ref{SD5micro}) once we take
into account that the latter represents indeed several coincident D5 branes.

Let us now analyze the meaning of the magnetic DBI field. In the CS piece of
the action for our D5 we would have a coupling to $C^{(2)}+C^{(0)}B_{NS}$ like

\begin{equation}
S_{CS}=\frac{T_5}{2}\int_{S^4}B\wedge B\int dt dr\
(C^{(2)}+C^{(0)}B_{NS})_{tr}\ .
\end{equation}

\noindent Therefore we see that the effect of adding these instantons on our
brane is to give it an extra D1 charge along the $t,r$ directions. From
this point of view, since on the D1 branes we also have an electric DBI field
which has the meaning of F1 strings along the $t,r$ directions, these kind of
configurations would describe dyonic strings. Since the macroscopic
configurations we are interested in just carry F-string charge and no D1
charge, we have to unplug it by setting the instanton number to zero, which in
turn means to eliminate the extra term in (\ref{SD5microFULL}). 

\section{Dyonic branes}

As we have just seen, in order to capture the Wilson line in the antisymmetric
product representation we have to unplug an extra D1 charge. These D1 are
extended along the same direction in which we have the F1, namely the $t,\
r$. Therefore, for the complete configuration, the charge carried by the
brane when pinching off on the loop at the boundary is not just F1 charge
(Wilson loop) but also D1 charge. This suggests that (\ref{SD5macro})
and its microscopical counterpart indeed capture a Wilson-'t Hooft line.

Also, we could perform an analog computation for the symmetric tensor product
case. Focusing for simplicity in the macroscopical description, we can
consider the D3 brane with both electric and magnetic DBI field, thus with F1
and D1 charge dissolved on the brane. Since we are at $g_s=1$, and the D3
brane is self S-dual, as expected the value of the ``routhian'' for such a
brane is 

\begin{equation}
\mathcal{R}=\int dt dr\ 2T_1\frac{L^4r^2}{y^4}\Big(\sqrt{1+y'^2}\sqrt{1+\frac{y^4}{4r^4L^4}(M^2+\frac{P^2}{T_1^2})+}-1\Big)\ ,
\end{equation}

\noindent which is invariant under the exchange of the two charges ($M$
(magnetic) $\leftrightarrow$ $\frac{P}{T_1}$ (electric)).

\section{Conclusions}

In \cite{DF} D3 branes with electric flux representing F strings where used to
compute certain Wilson loops. The
strategy was essentially to capture the all genus Wilson loop by means of a D3
brane since this D3 brane is assumed to be the effective description of the
multi-string configuration. Therefore it incorporates in a smooth manner all
the interactions between the strings, which amount to the subleading
corrections to the Wilson loop. 

Restricting ourselves to the Wilson (or Polyakov) lines,
which in the dual picture are represented by branes which pinch off at the
boundary on a line, as described in \cite{Y1}, \cite{Y2} and \cite{Gomis}, the
line in the symmetric tensor product representation can be computed by means
of a D3 brane carrying F1 strings, while the line in the antisymmetric product
can be computed with D5 with electric flux representing F1 strings.

For the case of the Wilson line in the symmetric product representation, since
F1 strings naturally polarize into D3 branes, it is to be expected that there
should exist a microscopical description in terms of fundamental strings
expanding due to dielectric effect to the D3, which in a suitable limit
overlaps with the effective description in terms of the D3 brane with
dissolved F1. We have explicitly shown this. Indeed, we have worked with the
S-dual picture, which captures the 't Hooft line. However, due to the manifest
S-duality invariance, the functional form will be the same for both the 't
Hooft and the Wilson line. Actually, it can be seen that an action for
dielectric F1 can be constructed by means of a chain of dualities starting
with the action of \cite{dielectricgravitons}, which however overlaps with
Myers action in this background. This is precisely a consequence of the
anticipated fact (see \cite{DF}) that both the Wilson and 't Hooft lines have
the same functional form.

The case of the Wilson line in the antisymmetric product representation is
somehow more involved, since naturally F1 strings do not expand to D5
branes. In turn, D1 strings expand naturally to D5 branes. Thus, we achieved a
microscopical description of such lines in terms of dielectric D1 strings
carrying the necessary electric flux, which expand to a fuzzy $S^4$. Once we
make sure that the total D1 charge is zero, we have a perfect agreement, again
in a suitable limit, with the effective description in terms of a D5 with
electric flux.

It is worth to note that, along the lines in \cite{CLT}, microscopically we
obtain the action for $n$ D5 rather that the action for a single D5. Thus, we
would be describing the multi-D5 configurations of \cite{Gomis}.

Inspired by the D5-brane case, in which we have the possibility of switching
an instantonic DBI field which represents D1 charge in addition to the desired
F1 charge, we can speculate on the possibility of capturing Wilson-'t Hooft
lines in terms of D3 (D5) branes for the symmetric (antisymmetric) tensor
product  representation. Since the charge pinching off at the boundary on the
loop is both F1 and D1, we expect that these branes capture some information
about the Wilson-'t Hooft line. It would be interesting to further study this
configurations.

In addition, this picture of the Wilson loop opens a new perspective, since in
principle it is possible to compute the loop at any charge using the work
in \cite{sanjaye} to compute at finite $N$ the symmetrized trace. This might
be more interesting for the circular Wilson loop, which is described by a
random matrix model. The circular Wilson loop calculated in \cite{ESZ}, breaks
another set of supersymetries as the one considered here. Since this loop can
be represented by means of a D3 with electric flux pinching off on the loop at
the boundary, in close parallelism with the Wilson line in the symmetric
product representation, we belive that a similar microscopical description
should be possible for that loop, which in turn could be related to the matrix
model to which the circular loop is dual.

\vspace{0.5cm}

\textbf{Acknowledgements}

\vspace{0.5cm}

It is a pleasure to thank B.Janssen, Y.Lozano and J.P.Resco for useful comments
on the manuscript. 

\vspace{0.5cm}

%

\appendix

\section{(No) D3-branes}

Let us analyze the configurations of D3 branes studied in
\cite{HK}. Motivated by \cite{DF}, and given (\ref{BHAdS}), in \cite{HK} it is
suggested that the thermal multiply wound Polyakov loop is represented by a D3
brane with worldvolume coordinates $\{t,\alpha,\Omega_2\}$, with $r=r(\alpha)$
and non-vanishing $F_{t\alpha}$. However, as shown in \cite{HK}, this
configuration does not behave as expected for such a loop. 

In addition to the metric background of (\ref{BHAdS}), there is a 4-form
potential. The relevant piece for the case at hand is

\begin{equation}
C^{(4)}=-iR^4r^4\sin^2\alpha\ dt\wedge d\alpha\wedge d\omega_2\ ,
\end{equation}

It is straightforward to check that the action for the system is

\begin{equation}
\label{SD3macro}
S=\int dt d\alpha\ 4\pi T_3R^4r^2\sin^2\alpha
\sqrt{r^2f+r'^2-\frac{F^2}{R^4}}-\int dt d\alpha\ 4\pi T_3R^4 r^4\sin^2\alpha\ .
\end{equation}

\noindent In (\ref{SD3macro}) we are setting $F_{t\alpha}=iF$, following
\cite{HK}. In addition, the prime stands for derivative with respect to
$\alpha$.

Since in (\ref{SD3macro}) the DBI potential does not appear, we can Legendre
transform to get a ``routhian'' in terms of a conserved momentum
$\frac{\partial L}{\partial F}=P$. It is easy to check that such a routhian
is 

\begin{equation}
\label{RD3macro}
\mathcal{R}_{Macro}=\int dt d\alpha\ T_1R^4\sqrt{r^2f+r'^2}\sqrt{\frac{P^2}{T_1^2R^4}+4r^4\sin^4\alpha}- \int dt d\alpha\ 2T_1R^4r^4\sin^2\alpha\ .
\end{equation}

\noindent In (\ref{RD3macro}) we have made use of the fact that in our
conventions $2\pi l_s^2=1$ to write the routhian in terms of $T_1$ for later
convenience.

In order to find a microscopical description of this configuration as in the
subsection it is convenient to go to cartesian coordinates

\begin{equation}
ds^2=R^2fdt^2+\frac{R^2}{f}dr^2+R^2r^2\big(d\alpha^2+\sin^2\alpha
d\vec{x}^2\big)+R^2d\Omega_5^ 2\ ,
\end{equation}

\noindent being $f$ the same as above.

In addition, the relevant piece of the RR potential is

\begin{equation}
C^{(4)}=-iR^4r^4\sin^2\alpha x^k\epsilon_{ijk}dt\wedge
d\alpha\wedge dx^i\wedge dx^j\ ,
\end{equation} 

Note that we are assuming that $\vec{x}^2=1$.

Given the macroscopic configuration, it is natural to assume that it can be
regarded as the effective theory of dielectric strings stretching along the
$\{t,\alpha\}$ directions, and expanding to the transverse 2-sphere whose
cartesian coordinates are the $\vec{x}$. In addition, we will assume that
$r=r(\alpha)$. 

Upon particularizing Myers action for our background we arrive to the
following action

\begin{equation}
\label{RD3micro}
\mathcal{R}_{micro}=\int dt d\alpha\ T_1R^4\Bigg(\sqrt{r^2f+r'^2}\sqrt{\frac{M^2}{R^4}+\frac{M^2}{C_2}4r^4\sin^4\alpha}-\frac{M}{\sqrt{C_2}}\frac{2r^4}{g_s}\sin^2\alpha\Bigg)\ .
\end{equation}

If we now take the Myers limit, in which (\ref{RD3macro}) and
(\ref{RD3micro})are expected to agree, we see that once we are in the large
$M$ limit, when $\frac{M}{\sqrt{C_2}}\rightarrow 1$, this is indeed the case,
and we have a perfect matching between both descriptions.

\end{document}